\documentclass[letterpaper,12pt]{article}
\usepackage[margin=1in]{geometry}
\usepackage{verbatim}
\usepackage{amsmath}
\usepackage{amssymb}
\usepackage{graphicx}
\usepackage{array}
\usepackage{amsthm}
\usepackage{accents}
\usepackage{subfig}
\usepackage[round,authoryear]{natbib}
\usepackage[breaklinks,hidelinks]{hyperref}

\def\logit{{\rm logit}}

\def\epsilon{\varepsilon}

\def\0s{{\bf 0}}

\newlength{\dhatheight}
\newcommand{\doublehat}[1]{{%
 \settoheight{\dhatheight}{\ensuremath{\hat{#1}}}%
 \addtolength{\dhatheight}{-0.2ex}%
 \hat{\vphantom{\rule{1pt}{\dhatheight}}%
 \smash{\hat{#1}}}%
}}

\newlength{\dtildeheight}
\newcommand{\doubletilde}[1]{{%
 \settoheight{\dtildeheight}{\ensuremath{\tilde{#1}}}%
 \addtolength{\dtildeheight}{-0.16ex}%
 \tilde{\vphantom{\rule{1pt}{\dtildeheight}}%
 \smash{\tilde{#1}}}%
}}

\newtheorem{theorem}{Theorem}[section]

\newtheorem{observe}[theorem]{Observation}
\newtheorem{remark1}[theorem]{Remark}

\newenvironment{remark}{\begin{remark1} \rm}{\end{remark1}}

\title{Testing goodness-of-fit for logistic regression}
\author{Mark Tygert and Rachel Ward}

\begin{document}

\maketitle

\begin{abstract}
Explicitly accounting for all applicable independent variables,
even when the model being tested does not, is critical in testing
goodness-of-fit for logistic regression.
This can increase statistical power by orders of magnitude.
\end{abstract}

\section{Introduction}
\label{intro}

Testing goodness-of-fit for logistic regression has received
a remarkable amount of attention, with contributions
from~\cite{hosmer-lemeshow1}, \cite{berran-millar}, \cite{su-wei},
\cite{royston}, \cite{stute-zhu}, \cite{wlthm}, \cite{pan-lin},
and~\cite{allison}, among many others;
see, for example, the references of~\cite{hosmer-lemeshow2}.
The discussion in the present paper is closely related.
However, even when computing exact P-values as in earlier works,
as summarized in the appendix below,
omissions in the standard tests can reduce the standards' statistical power
by orders of magnitude.
The present, introductory section addresses the omissions.
For details, see Section~\ref{examples} or the concluding remarks
(Section~\ref{conclusion}).

A representative use of logistic regression is to model the absence (0)
or presence (1) of coronary heart disease
--- resulting in a ``dependent variable'' for coronary heart disease that is
binary/dichotomous (meaning that the observed values are zeros and ones).
The regression predicts the dependent variable via ``independent variables''
such as age, cholesterol level, diastolic and systolic blood pressure,
and others listed in Section~\ref{introex} below.

To be precise, the null hypothesis for a logistic regression
of a binary/dichotomous random variable $Y_k$ on real numbers $x_{j,k}$
(with $k = 1$,~$2$, \dots, $n$, and $j = 1$,~$2$, \dots,
$m\hspace{.38em}{^{_{_{\scriptstyle{\ll}}}}}\hspace{.25em}n$),
given observations $y_1$,~$y_2$, \dots, $y_n$
of $Y_1$,~$Y_2$, \dots, $Y_n$, respectively, is
\begin{multline}
\label{GLMhyp}
H_0 : y_1, y_2, \dots, y_n
  \hbox{ are draws from independent Bernoulli distributions}\\
  \hbox{with means }
  \hat{\mu}_1, \hat{\mu}_2, \dots, \hat{\mu}_n,\hbox{ respectively,}
\end{multline}
\begin{equation}
\label{GLMaux}
\logit(\hat{\mu}_k) = \hat\beta^{(0)} + \sum_{j=1}^l \hat\beta^{(j)} x_{j,k}
\end{equation}
for $k = 1$,~$2$, \dots, $n$,
where the vector $\beta$ is a nuisance parameter,
$\hat\beta$ is its maximum-likelihood estimate, $l$ is a nonnegative integer
no greater than $m$, and
\begin{equation}
\logit(p) = \ln\left(\frac{p}{1-p}\right)
\end{equation}
is the natural logarithm of the odds $p/(1-p)$.
Beware that interpreting this type of null hypothesis
involves subtleties explicated by~\cite{perkins-tygert-ward4} and others.
In this general formulation, the number $l$ of terms in the sum
from~(\ref{GLMaux}) may be less than the total number $m$
of independent variables (while $m$ itself should be
significantly less than $n$ to avoid overfitting);
nonetheless, accounting for all applicable
independent variables when testing goodness-of-fit is very important,
as discussed shortly and further detailed below in Section~\ref{exintro}.

Goodness of fit gauges the consistency of the given observed values
with the model assumed under the null hypothesis.
The observed values $y_1$,~$y_2$, \dots, $y_n$ are zeros and ones,
so probing their distribution requires aggregating or accumulating
these very low counts, just as in the case of draws
from continuous probability densities
(the continuous case requires use of cumulative distribution functions,
probability density estimation, Neyman smooth tests,
or something similarly aggregative).
A cumulative measure of the distance between the observed data
and the model assumed under the null hypothesis
is the discrete Kolmogorov-Smirnov statistic
(recommended by~\cite{horn} and many others), namely,
\begin{equation}
\label{GLMKS}
d
= \max_{1 \le j \le n} \left| \sum_{k=1}^j y_{\sigma_k}
                            - \sum_{k=1}^j \hat{\mu}_{\sigma_k} \right|
= \max_{1 \le j \le n} \left| \sum_{k=1}^j r_{\sigma_k} \right|,
\end{equation}
where the ordering $\sigma$ is a permutation
of the integers $1$,~$2$, \dots, $n$, and the residual $r_k$ is
\begin{equation}
r_k = y_k-\hat{\mu}_k
\end{equation}
for $k = 1$,~$2$, \dots, $n$.
Please note that $\hat{\mu}_1$,~$\hat{\mu}_2$, \dots, $\hat{\mu}_n$
are from~(\ref{GLMaux}); (\ref{GLMaux}) is similar to~(\ref{GLMaux2}) below,
but (\ref{GLMaux2}) influences only the permutation $\sigma$.

Critically, the ordering $\sigma$ in~(\ref{GLMKS}) should take into account
all applicable independent variables, not just those included in the model
of~(\ref{GLMhyp}) and~(\ref{GLMaux}).
If a suitable ordering for the observations is not known a priori,
there is a good substitute, namely to choose a permutation $\sigma$
of the integers $1$,~$2$, \dots, $n$ such that
$\tilde{\mu}_{\sigma_1} \le \tilde{\mu}_{\sigma_2} \le \dots
\le \tilde{\mu}_{\sigma_n}$, replacing~(\ref{GLMaux}) with a fit
to all the data via
\begin{equation}
\label{GLMaux2}
\logit(\tilde{\mu}_k) = \tilde\beta^{(0)}
                      + \sum_{j=1}^m \tilde\beta^{(j)} x_{j,k}
\end{equation}
for $k = 1$,~$2$, \dots, $n$ (notice that the sum here runs up to $m$,
not just $l$), where $\tilde{\mu}_1$, $\tilde{\mu}_2$, \dots, $\tilde{\mu}_n$
are the estimated means of the postulated independent Bernoulli distributions.
That is, this natural ordering sorts based on the estimated means
$\tilde{\mu}_1$, $\tilde{\mu}_2$, \dots, $\tilde{\mu}_n$
when estimated using all applicable independent variables,
not using only the $l$ terms appearing in the sum from~(\ref{GLMaux}).
This ordering incorporates information
from all values $x_{1,1}$,~$x_{1,2}$, \dots, $x_{m,n}$,
as these values influence the estimated values
$\tilde{\mu}_1$,~$\tilde{\mu}_2$, \dots, $\tilde{\mu}_n$ in~(\ref{GLMaux2}).

Goodness of fit is gauged via the ``P-value.'' As reviewed in the appendix,
the P-value associated with $d$ defined in~(\ref{GLMKS})
is the proportion of Monte-Carlo simulations for which the distance $d$
is greater than or equal to the distance $d$ for the original, observed data,
in the limit of a large number of simulations,
each simulated according to~(\ref{GLMhyp}) and~(\ref{GLMaux}).
If a P-value is very small, then we can be confident that
the model, given in~(\ref{GLMhyp}) and~(\ref{GLMaux}), is not consistent
with the data, even when allowing for the expected statistical fluctuations.

The next section illustrates that the above approach can be very powerful.
This introductory section will now conclude with several incidental remarks
about other approaches:

\begin{remark}
If we replace the natural ordering (discussed above)
with a permutation $\sigma$ satisfying
$r_{\sigma_1} \le r_{\sigma_2} \le \dots \le r_{\sigma_n}$,
then (\ref{GLMKS}) simplifies to
\begin{equation}
\label{GLMmax}
d = \frac{1}{2} \sum_{k=1}^n |r_k|.
\end{equation}
Maximizing~(\ref{GLMKS}) over all permutations $\sigma$
of the integers $1$,~$2$, \dots, $n$ also yields~(\ref{GLMmax}).
The simplification to~(\ref{GLMmax}) is due to~\cite{hoeffding}
(see the top of page 396 in Hoeffding's article);
this follows from the fact that the sum of the residuals is 0, that is,
$\sum_{k=1}^n r_k = 0$ (the sum of the residuals is 0
on account of the constant term $\hat\beta^{(0)}$ in~(\ref{GLMaux})).
\end{remark}

\begin{remark}
An often appealing alternative to the Kolmogorov-Smirnov statistic defined
in~(\ref{GLMKS}) is the Kuiper statistic
\begin{equation}
\label{GLMK}
d
= \max_{1 \le j \le n} \sum_{k=1}^j r_{\sigma_k}
- \min_{1 \le j \le n} \sum_{k=1}^j r_{\sigma_k}.
\end{equation}
Under the ordering for which the permutation $\sigma$ satisfies
$r_{\sigma_1} \le r_{\sigma_2} \le \dots \le r_{\sigma_n}$,
(\ref{GLMK}) simplifies to~(\ref{GLMKS}) and hence to~(\ref{GLMmax}),
since $\sum_{k=1}^n r_{\sigma_k} = \sum_{k=1}^n r_k = 0$,
while this choice of ordering ensures
$\sum_{k=1}^j r_{\sigma_k} \le 0$ for all $j=1$,~$2$, \dots, $n$.
In fact, (\ref{GLMK}) simplifies to~(\ref{GLMKS}) and hence to~(\ref{GLMmax}),
for any permutation $\sigma$ such that
$r_{\sigma_{k+1}} \le r_{\sigma_{k+2}} \le \dots \le r_{\sigma_n}
\le r_{\sigma_1} \le r_{\sigma_2} \le \dots \le r_{\sigma_k}$
for some positive integer $k$.
Indeed, the value of (\ref{GLMK}) for such a permutation
is the same as for a permutation satisfying 
$r_{\sigma_1} \le r_{\sigma_2} \le \dots \le r_{\sigma_n}$;
this invariance is the principal appeal of the Kuiper statistic,
as discussed by~\cite{stephens2} and Section~14.3.4
of~\cite{press-teukolsky-vetterling-flannery}.
\end{remark}

\begin{remark}
In this remark, we consider the case
when the vectors $(x_{1,k}, x_{2,k}, \dots, x_{l,k})$ are different
for different values of $k$.
This is the case for the above example of modeling the absence or presence
of coronary heart disease, provided that no two subjects have exactly the same
age, cholesterol level, diastolic and systolic blood pressure, and so on.
The deviance (also known as the log--likelihood ratio or ``$G^2$'') is then
\begin{equation}
\label{GLM1}
g^2 = -2 \sum_{k=1}^n \Bigl( y_k \ln(\hat{\mu}_k)
                           + (1-y_k) \ln(1-\hat{\mu}_k) \Bigr)
= -2 \sum_{y_k=0} \ln(1-\hat{\mu}_k) - 2 \sum_{y_k=1} \ln(\hat{\mu}_k),
\end{equation}
where $\hat{\mu}_k$ is the maximum-likelihood estimate of the mean
for the Bernoulli distribution producing $y_k$ under the model;
$\hat{\mu}_k$ is defined in~(\ref{GLMaux}).
If $\hat{\mu}_k$ is small whenever $y_k=0$
and $1-\hat{\mu}_k$ is small whenever $y_k=1$ (that is,
the absolute residual $|r_k| = |y_k-\hat{\mu}_k|$ is small
for $k=1$,~$2$, \dots,~$n$), then
\begin{equation}
\label{GLM2}
-\sum_{y_k=0} \ln(1-\hat{\mu}_k) - \sum_{y_k=1} \ln(\hat{\mu}_k)
\approx \sum_{y_k=0} \hat{\mu}_k + \sum_{y_k=1} (1-\hat{\mu}_k).
\end{equation}
Moreover,
\begin{equation}
\label{GLM3}
\sum_{y_k=0} \hat{\mu}_k + \sum_{y_k=1} (1-\hat{\mu}_k)
= -\sum_{y_k=0} r_k + \sum_{y_k=1} r_k
= \sum_{y_k=0} |r_k| + \sum_{y_k=1} |r_k|
= \sum_{k=1}^n |r_k|,
\end{equation}
where $r_k = y_k-\hat{\mu}_k$ is the residual.
Combining~(\ref{GLM1}), (\ref{GLM2}), and~(\ref{GLM3}) yields that
\begin{equation}
\label{GLM4}
g^2 \approx 2 \sum_{k=1}^n |r_k|
\end{equation}
if the absolute residuals $|r_1|$,~$|r_2|$, \dots, $|r_n|$ are small.
Please note that the right-hand side of~(\ref{GLM4})
is exactly four times the right-hand side of~(\ref{GLMmax}).
Beware that the deviance $g^2$ is not terribly helpful for gauging
goodness-of-fit with logistic regression,
as the deviance is a deterministic function
of the estimated means of the postulated Bernoulli distributions
--- the deviance depends on the observed data only through
dependence on the estimated means.
In fact, \cite{mccullagh-nelder} and others have shown that
\begin{equation}
\label{GLMbaddev}
g^2
= 2 \sum_{k=1}^n \hat{\mu}_k \ln\left(\frac{1-\hat{\mu}_k}{\hat{\mu}_k}\right)
- 2 \sum_{k=1}^n \ln(1-\hat{\mu}_k)
\end{equation}
when the vectors $(x_{1,k}, x_{2,k}, \dots, x_{l,k})$,
with $k=1$,~$2$, \dots, $n$, are distinct;
please notice that the observations $y_1$,~$y_2$, \dots, $y_n$ do not appear
explicitly in the right-hand side of~(\ref{GLMbaddev})
(though of course the estimated means
$\hat{\mu}_1$,~$\hat{\mu}_2$, \dots, $\hat{\mu}_n$ do depend
on the observations $y_1$,~$y_2$, \dots, $y_n$).
\end{remark}

\begin{remark}
\label{hlrem}
The tests of~\cite{hosmer-lemeshow1,hosmer-lemeshow2}
replace the Kolmogorov-Smirnov statistic $d$ defined in~(\ref{GLMKS})
with $\chi^2$ for the model of binomial distributions
corresponding to quantiles (typically the deciles)
for the estimated means of the postulated Bernoulli distributions.
For example, if $n$ is divisible by $10$ (for notational convenience),
then we could replace $d$ defined in~(\ref{GLMKS}) with
\begin{equation}
\label{hoslem}
h_l = \sum_{k=1}^{10} \frac{(n_k-\hat{\eta}_k)^2}
                           {\hat{\eta}_k (1-10\hat{\eta}_k/n)},
\end{equation}
where $n_k$ is the sum of the observed values $y_1$,~$y_2$, \dots, $y_n$
from the $k$th group
(that is, $n_k$ is the number of values from the $k$th group
that are equal to 1),
grouping according to the deciles of the estimated means
$\hat{\mu}_1$,~$\hat{\mu}_2$, \dots, $\hat{\mu}_n$,
and where $\hat{\eta}_k$ is the sum of the estimated means
$\hat{\mu}_1$,~$\hat{\mu}_2$, \dots, $\hat{\mu}_n$ in the $k$th decile
(that is, in the $k$th group). Notice that the denominator in~(\ref{hoslem})
would be the expected value of the numerator if $n_k$ were drawn
from a binomial distribution whose maximal possible value is $n/10$
and whose mean is $\hat{\eta}_k$ (conditional on knowing $\hat{\eta}_k$).
This approach of~\cite{hosmer-lemeshow1,hosmer-lemeshow2} is similar
to the cumulative (Kolmogorov-Smirnov) approach detailed
in the present section, but involves explicit binning.
\cite{allison} criticizes the binning
and provides references to many other critiques.
\end{remark}

\section{Examples}
\label{examples}

\subsection{Overview}
\label{exintro}

This subsection introduces two thought experiments,
as well as their illustration below
via the computer-assisted analysis of three real data sets.
The statistic detailed in Section~\ref{intro}
--- Kolmogorov-Smirnov ordered based on $\tilde{\mu}$ ---
far outperforms the standard statistic described in Remark~\ref{hlrem}
--- Hosmer-Lemeshow ordered based on $\hat{\mu}$.
Whereas $\hat{\mu}$ considers only the $l$ summands
in~(\ref{GLMaux}), $\tilde{\mu}$ takes into account
all $m$ independent variables in~(\ref{GLMaux2}).

Beware that accounting for all applicable independent variables
when testing goodness-of-fit is essential;
using only the independent variables included
in~(\ref{GLMaux}) is not always sufficient. Consider, for example,
the case when $l = 0$ in~(\ref{GLMaux}), that is,
when the sum in~(\ref{GLMaux}) is absent.
The resulting logistic regression amounts to regressing
$y_1$,~$y_2$, \dots, $y_n$ against a constant;
the estimated means in~(\ref{GLMaux}) are then all the same,
that is, $\hat{\mu}_1 = \hat{\mu}_2 = \dots = \hat{\mu}_n$.
Unless nearly all $y_1$,~$y_2$,~\dots,~$y_n$ are equal,
the fit cannot possibly be good, yet no standard test for goodness of fit
can detect the poor fit.
Indeed, any ordering for the observations must be random
with the standard tests, since the estimated means are all the same,
$\hat{\mu}_1 = \hat{\mu}_2 = \dots = \hat{\mu}_n$.
The standard tests consider only the independent variables
appearing in~(\ref{GLMaux}) --- surely not enough when $l=0$.
Without substantial information from independent variables
to inform the ordering and the associated accumulation or aggregation,
the data cannot possibly invalidate the tested model.
The values of $y_1$,~$y_2$,~\dots,~$y_n$ (each 0 or 1) are not informative
if their ordering is totally random.
The data can invalidate the tested model only when
there are variables that can inform the ordering,
possibly including variables not included in the tested model.

Similarly, if $l = 1$ and the values $x_{1,1}$,~$x_{1,2}$, \dots, $x_{1,n}$
are drawn independently and with identical distributions (i.i.d.)\
from the uniform distribution over $(0,1)$,
with no relation to $y_1$,~$y_2$, \dots, $y_n$,
then the fit via~(\ref{GLMaux}) cannot possibly be good.
Yet, all standard tests for goodness of fit (which consider only those
independent variables appearing in~(\ref{GLMaux}), not any ``extras''
from the rest of the data set) cannot detect the poor fit.
Moreover, such problems are not limited to thought experiments,
as \cite{hhll}, \cite{wlthm}, and the remainder of the present paper
illustrate via the analysis of real data.

The following examples analyze several data sets,
referring to the test statistic detailed
in Section~\ref{intro} as Kolmogorov-Smirnov ordered based on $\tilde{\mu}$,
and we recommend Kolmogorov-Smirnov ordered based on $\tilde{\mu}$
for general-purpose use;
the standard statistic (detailed in Remark~\ref{hlrem})
is Hosmer-Lemeshow grouped based on $\hat{\mu}$
(not $\tilde{\mu}$), which performs relatively poorly in our tests.
The other statistics mentioned in the tables below (namely, $G^2$,
Freeman-Tukey, $\chi^2$, and the Euclidean distance) are not really relevant
to testing goodness-of-fit for logistic regression,
as they do not depend on the ordering;
the tables include their P-values only for completeness,
illustrating their low statistical power.
As always, if any of the P-values is very small, then we can have confidence
that the model in~(\ref{GLMhyp}) and~(\ref{GLMaux}) does not yield a good fit;
that is, we can have confidence that the observed data is not consistent
(up to the expected statistical fluctuations)
with assuming (\ref{GLMhyp}) and~(\ref{GLMaux}).

We used ``glmfit'' in Matlab for all calculations,
incorporating the appendix to compute all P-values;
Remark~\ref{1error-bars} discusses the resulting accuracy.
We calculated each P-value via $i =$ 4,000,000 Monte-Carlo simulations;
running so many simulations is not really necessary.

\subsection{Data from~\texorpdfstring{\cite{finney}}{Finney}}

Table~\ref{finneytab} displays the P-values for several goodness-of-fit tests
applied to the classic data set of~\cite{finney}.
Table~\ref{finneydat} displays the data set,
which consists of $n=39$ observations
of a dependent variable together with 2 independent variables.

Table~\ref{finneytab} reports on four experiments.
First, we set $l=0$, i.e., omit the sum in~(\ref{GLMaux}) entirely,
while retaining $m=2$ in~(\ref{GLMaux2}).
Second, we set $l=1$ and generate a new, additional independent variable,
drawing $x_{1,1}$,~$x_{1,2}$, \dots, $x_{1,n}$ i.i.d.\
from the uniform distribution over $(0,1)$.
Third, we discard the extra random independent variable,
retaining both original independent variables,
with $l=2$ and $m=2$.
Fourth, we again include the extra random independent variable
(making $m=2+1$), but do not include it in the tested model (so~$l=2$).

For the Hosmer-Lemeshow statistics of Remark~\ref{hlrem},
we consider different groupings.
In the first, with 3 groups in all, each group contains 13 observations.
In the second, with 5 groups in all,
the initial 4 groups contain 8 observations each, and the last contains 7.

\subsection{Data from~\texorpdfstring{\cite{hosmer-lemeshow2}}
                                           {Hosmer and Lemeshow}}

Table~\ref{uistab} displays the P-values for several goodness-of-fit tests
applied to the ``UIS'' data set of \cite{hosmer-lemeshow2}.
This data set consists of $n=575$ observations
of a dependent variable ``dfree'' together with 11 independent variables
``age,'' ``beck,'' ``ndrgfp1,'' ``ndrgfp2,'' ``ivhx\_2,'' ``ivhx\_3,''
``race,'' ``treat,'' ``site,'' ``ageXndrgfp1,'' and ``raceXsite''
(these include transformations and products
of the original variables ``dfree,'' ``age,'' ``beck,'' ``ndrugtx,'' ``ivhx,''
``race,'' ``treat,'' and ``site,'' as detailed in Chapter~4
of~\cite{hosmer-lemeshow2}).
For the Hosmer-Lemeshow statistics of Remark~\ref{hlrem},
we aggregate the data into 10 groups,
with the initial 9 containing 58 observations each, and the last containing 53.

Table~\ref{uistab} reports on five experiments.
First, we set $l=0$, i.e., omit the sum in~(\ref{GLMaux}) entirely,
while retaining $m=11$ in~(\ref{GLMaux2}).
Second, we set $l=1$ and generate a new, additional independent variable,
drawing $x_{1,1}$,~$x_{1,2}$, \dots, $x_{1,n}$ i.i.d.\
from the uniform distribution over $(0,1)$.
Third, we discard the extra random independent variable,
retaining the original $m=11$, and set $l=9$,
taking the independent variables in the regression to be
``age,'' ``beck,'' ``ndrgfp1,'' ``ndrgfp2,'' ``ivhx\_2,'' ``ivhx\_3,''
``race,'' ``treat,'' and ``site''
(these are all those not involving products of other variables).
Fourth, we include all 11 original independent variables,
with both $l=11$ and $m=11$.
Fifth, we again include the extra random independent variable
(making $m=11+1$), but do not include it in the tested model (so $l=11$).

\subsection{Data from~\texorpdfstring{\cite{kleinbaum-klein}}
                                           {Kleinbaum and Klein}}
\label{introex}

Table~\ref{kktab} displays the P-values for several goodness-of-fit tests
applied to the ``Evans County'' data set of~\cite{kleinbaum-klein}.
This data set consists of $n=609$ observations
of a dependent variable ``chd'' together with 10 independent variables
``age,'' ``cat,'' ``chl,'' ``dbp,'' ``ecg,'' ``hpt,'' ``sbp,'' ``smk,''
``catXchl,'' and ``catXhpt''
(these include products of the original variables ``chd,'' ``age,'' ``cat,''
``chl,'' ``dbp,'' ``ecg,'' ``hpt,'' ``sbp,'' and ``smk,''
as in model ``EC4'' from Chapter~9 of~\cite{kleinbaum-klein}).
For the Hosmer-Lemeshow statistics of Remark~\ref{hlrem},
we aggregate the data into 10 groups,
with the initial 9 containing 61 observations each, and the last containing 60.

Table~\ref{kktab} reports the results of five experiments.
First, we set $l=0$, i.e., omit the sum in~(\ref{GLMaux}) entirely,
while retaining $m=10$ in~(\ref{GLMaux2}).
Second, we set $l=1$ and generate a new, additional independent variable,
drawing $x_{1,1}$,~$x_{1,2}$, \dots, $x_{1,n}$ i.i.d.\
from the uniform distribution over $(0,1)$.
Third, we discard the extra random independent variable,
retaining the original $m=10$, and set $l=6$,
taking the independent variables in the regression to be
``age,'' ``cat,'' ``chl,'' ``ecg,'' ``hpt,'' and ``smk''
--- those included for model ``EC3'' in Chapter~9 of~\cite{kleinbaum-klein}.
Fourth, we include all 10 original independent variables,
with both $l=10$ and $m=10$.
Fifth, we again include the extra random independent variable
(making $m=10+1$), but do not include it in the tested model (so $l=10$).

\begin{table}
\caption{$x_{1,k}$, $x_{2,k}$, and~$y_k$ for $k=1$,~$2$, \dots, $39$,
         from~\cite{finney}; \cite{finney} refers to $y$ as ``response''}
\label{finneydat}
\begin{center}
\begin{tabular}{rrr}
$x_1$ & $x_2$ & $y$
\\\hline
1.57 & 0.92 & 1 \\
1.54 & 1.04 & 1 \\
1.10 & 1.40 & 1 \\
0.88 & 1.18 & 1 \\
0.90 & 1.51 & 1 \\
0.85 & 1.54 & 1 \\
0.78 & 0.88 & 0 \\
1.04 & 1.23 & 0 \\
0.95 & 0.88 & 0 \\
0.95 & 0.65 & 0 \\
0.90 & 0.76 & 0 \\
0.74 & 1.44 & 0 \\
0.78 & 1.48 & 0 \\
1.15 & 1.37 & 1 \\
0.88 & 1.57 & 1 \\
1.36 & 1.21 & 1 \\
1.51 & 1.20 & 1 \\
0.93 & 1.15 & 1 \\
1.23 & 1.03 & 0 \\
1.26 & 1.26 & 1 \\
0.60 & 1.30 & 0 \\
0.98 & 1.13 & 0 \\
1.13 & 1.13 & 0 \\
1.18 & 1.13 & 0 \\
1.20 & 1.25 & 1 \\
0.78 & 1.18 & 0 \\
1.26 & 1.18 & 1 \\
0.98 & 1.28 & 0 \\
1.28 & 0.98 & 1 \\
1.20 & 0.60 & 0 \\
1.43 & 0.88 & 1 \\
1.37 & 0.48 & 0 \\
1.04 & 1.26 & 0 \\
1.04 & 1.34 & 1 \\
1.08 & 1.30 & 1 \\
0.90 & 1.52 & 1 \\
0.98 & 1.28 & 0 \\
0.88 & 1.28 & 0 \\
1.11 & 1.21 & 1
\end{tabular}
\end{center}
\end{table}

\begin{table}
\caption{P-values for the data set of~\cite{finney}
(see Table~\ref{finneydat}); $n=39$}
\label{finneytab}
\renewcommand{\arraystretch}{1.5}
\setlength{\extrarowheight}{0em}
\setcounter{footnote}{0}
\begin{center}
\begin{tabular}{m{1.61in}llll}
& \parbox{2.9em}{$l=0$;\footnotemark\\$m=2$} & \parbox{4.0em}{$l=1$;\footnotemark\\$m=2+1$} & \parbox{2.9em}{$l=2$;\footnotemark\\$m=2$} & \parbox{4.0em}{$l=2$;\footnotemark\\$m=2+1$} \\[1em]\hline
Kolmogorov-Smirnov (ordering based on $\tilde{\mu}$) & $.0000003^\dagger$ & $.0000003^\dagger$ & .0075 & .039 \\
Kolmogorov-Smirnov (ordering based on $\hat{\mu}$) & .249 & .452 & .0075 & .0075 \\
Kolmogorov-Smirnov (ordering based on $r$) & .134 & .053 & .355 & .355 \\
$G^2$ (the deviance, log--likelihood-ratio, \dots) & .248 & .053 & .324 & .324 \\
Freeman-Tukey (Hellinger distance) & .517 & .239 & .250 & .250 \\
$\chi^2$ (sum of the squares of Pearson residuals) & .385 & .405 & .182 & .182 \\
Euclidean-distance (unweighted $\chi^2$) & .236 & .053 & .393 & .393 \\
Hosmer-Lemeshow (with 3 groups from $\tilde{\mu}$) & .582 & .248 & .107 & .065 \\
Hosmer-Lemeshow (with 3 groups from $\hat{\mu}$) & .590 & .298 & .107 & .107 \\
Hosmer-Lemeshow (with 5 groups from $\tilde{\mu}$) & .658 & .324 & .787 & .695 \\
Hosmer-Lemeshow (with 5 groups from $\hat{\mu}$) & .597 & .201 & .787 & .787
\end{tabular}
\end{center}
\setcounter{footnote}{0}
\footnotemark{}i.e., omitting the sum in~(\ref{GLMaux}) entirely,
while retaining the original $m=2$

\footnotemark{}i.e., with an extra independent variable,
with $x_{1,1}$,~$x_{1,2}$, \dots, $x_{1,n}$ drawn i.i.d.\ from $U(0,1)$

\footnotemark{}i.e., including both original independent variables

\footnotemark{}i.e., with an extra independent variable,
with $x_{3,1}$,~$x_{3,2}$, \dots, $x_{3,n}$ drawn i.i.d.\ from $U(0,1)$,
but without including the extra variable in the model being tested
(while still including both original independent variables)

$^\dagger$in these cases,
only 1 simulation (out of 4,000,000) produced a Kolmogorov-Smirnov statistic
at least as large as that for the original, observed data

\end{table}

\begin{table}
\caption{P-values for the ``UIS'' data set of~\cite{hosmer-lemeshow2}; $n=575$}
\label{uistab}
\renewcommand{\arraystretch}{1.5}
\setlength{\extrarowheight}{0em}
\setcounter{footnote}{0}
\begin{center}
\begin{tabular}{m{1.6in}lllll}
& \parbox{3.3em}{$l=0$;\footnotemark\\$m=11$} & \parbox{4.5em}{$l=1$;\footnotemark\\$m=11+1$} & \parbox{3.3em}{$l=9$;\footnotemark\\$m=11$} & \parbox{3.4em}{$l=11$;\footnotemark\\$m=11$} & \parbox{4.5em}{$l=11$;\footnotemark\\$m=11+1$} \\[1em]\hline
Kolmogorov-Smirnov (ordering based on $\tilde{\mu}$) & .00001 & .00001 & .0049 & .736 & .861 \\
Kolmogorov-Smirnov (ordering based on $\hat{\mu}$) & .609 & .276 & .115 & .736 & .736 \\
Kolmogorov-Smirnov (ordering based on $r$) & .486 & .484 & .334 & .319 & .319 \\
$G^2$ (the deviance, log--likelihood-ratio, \dots) & .516 & .482 & .343 & .311 & .311 \\
Freeman-Tukey (Hellinger distance) & .516 & .478 & .314 & .286 & .286 \\
$\chi^2$ (sum of the squares of Pearson residuals) & .473 & .734 & .740 & .300 & .300 \\
Euclidean-distance (unweighted $\chi^2$) & .507 & .484 & .317 & .319 & .319 \\
Hosmer-Lemeshow (for deciles of $\tilde{\mu}$) & .023 & .017 & .991 & .673 & .594 \\
Hosmer-Lemeshow (for deciles of $\hat{\mu}$) & .688 & .249 & .781 & .673 & .673
\end{tabular}
\end{center}
\setcounter{footnote}{0}
\footnotemark{}i.e., omitting the sum in~(\ref{GLMaux}) entirely,
while retaining the original $m=11$

\footnotemark{}i.e., with an extra independent variable,
with $x_{1,1}$,~$x_{1,2}$, \dots, $x_{1,n}$ drawn i.i.d.\ from $U(0,1)$

\footnotemark{}i.e., taking the independent variables in the regression to be
``age,'' ``beck,'' ``ndrgfp1,'' ``ndrgfp2,'' ``ivhx\_2,'' ``ivhx\_3,''
``race,'' ``treat,'' and ``site''
(these are all those not involving products of other variables),
while retaining the original $m=11$

\footnotemark{}i.e., including all 11 original independent variables

\footnotemark{}i.e., with an extra independent variable,
with $x_{12,1}$,~$x_{12,2}$, \dots, $x_{12,n}$ drawn i.i.d.\ from $U(0,1)$,
but without including the extra variable in the model being tested
(while still including all 11 original independent variables)

\end{table}

\begin{table}
\caption{P-values for the ``Evans County'' data set of~\cite{kleinbaum-klein};
$n=609$}
\label{kktab}
\renewcommand{\arraystretch}{1.5}
\setlength{\extrarowheight}{0em}
\setcounter{footnote}{0}
\begin{center}
\begin{tabular}{m{1.6in}lllll}
& \parbox{3.3em}{$l=0$;\footnotemark\\$m=10$} & \parbox{4.5em}{$l=1$;\footnotemark\\$m=10+1$} & \parbox{3.3em}{$l=6$;\footnotemark\\$m=10$} & \parbox{3.4em}{$l=10$;\footnotemark\\$m=10$} & \parbox{4.5em}{$l=10$;\footnotemark\\$m=10+1$} \\[1em]\hline
Kolmogorov-Smirnov (ordering based on $\tilde{\mu}$) & $\le$\,$.0000003^\dagger$ & $\le$\,$.0000003^\dagger$ & $\le$\,$.0000003^\dagger$ & .193 & .328 \\
Kolmogorov-Smirnov (ordering based on $\hat{\mu}$) & .357 & .905 & .738 & .193 & .193 \\
Kolmogorov-Smirnov (ordering based on $r$) & .471 & .474 & .431 & .418 & .418 \\
$G^2$ (the deviance, log--likelihood-ratio, \dots) & .519 & .472 & .412 & .357 & .357 \\
Freeman-Tukey (Hellinger distance) & .485 & .472 & .404 & .405 & .405 \\
$\chi^2$ (sum of the squares of Pearson residuals) & .354 & .427 & .759 & .010 & .010 \\
Euclidean-distance (unweighted $\chi^2$) & .514 & .474 & .431 & .451 & .451 \\
Hosmer-Lemeshow (for deciles of $\tilde{\mu}$) & $\le$\,$.0000003^\dagger$ & .000002 & .995 & .237 & .186 \\
Hosmer-Lemeshow (for deciles of $\hat{\mu}$) & .651 & .585 & .822 & .237 & .237
\end{tabular}
\end{center}
\setcounter{footnote}{0}
\footnotemark{}i.e., omitting the sum in~(\ref{GLMaux}) entirely,
while retaining the original $m=10$

\footnotemark{}i.e., with an extra independent variable,
with $x_{1,1}$,~$x_{1,2}$, \dots, $x_{1,n}$ drawn i.i.d.\ from $U(0,1)$

\footnotemark{}i.e., taking the independent variables in the regression to be
``age,'' ``cat,'' ``chl,'' ``ecg,'' ``hpt,'' and ``smk''
--- those included for model ``EC3'' in Chapter~9 of~\cite{kleinbaum-klein}
--- while retaining the original $m=10$

\footnotemark{}i.e., including all 10 original independent variables

\footnotemark{}i.e., with an extra independent variable,
with $x_{11,1}$,~$x_{11,2}$, \dots, $x_{11,n}$ drawn i.i.d.\ from $U(0,1)$,
but without including the extra variable in the model being tested
(while still including all 10 original independent variables)

$^\dagger$in these cases,
no simulations (out of 4,000,000) produced a test statistic
at least as large as that for the original, observed data

\end{table}

\newpage

\section{Conclusion}
\label{conclusion}

The discrete Kolmogorov-Smirnov test with an ordering based
on all applicable independent variables produces P-values that are orders
of magnitude better
than those for the standards (such as the usual Hosmer-Lemeshow test)
in many circumstances for which the model clearly fits very poorly.
In particular, this happens if the model omits
significant explanatory variables that are in the given data.
The Kolmogorov-Smirnov approach is not the only possibility,
but the above examples (both the thought experiments
and the real data analyses) argue strongly in favor
of aggregating based on all applicable independent variables,
not based on just those incorporated into the model being tested.
In fact, this is a strong argument relevant to testing goodness-of-fit
for any regression with low counts, including the simplest logistic regression
that is the focus of the present paper.

\bigskip

\appendix
\section{Appendix: Computation of P-values}
\label{1computation}

This appendix describes Monte-Carlo simulations yielding estimates
and confidence intervals for P-values.
The standard errors of the estimates are inversely proportional
to the square root of the number of simulations;
the P-values being estimated are exact for any number $n$ of observations
and also have desirable properties in the limit that $n$ is large,
as detailed by \cite{perkins-tygert-ward4},
based on work of~\cite{romano}, \cite{henze}, \cite{cox}, and others.
To calculate a P-value, we first estimate $\beta$ from the given observations,
obtaining $\hat\beta$ in~(\ref{GLMaux}) and $\tilde\beta$ in~(\ref{GLMaux2}),
and then calculate the test statistic (such as $d$ in~(\ref{GLMKS})).
We next run many simulations.
To conduct a single simulation, we perform the following three-step procedure:
\begin{enumerate}
\item we generate $n$ independent draws according
      to~(\ref{GLMhyp}) and~(\ref{GLMaux}),
\item we fit the parameter $\beta$ from the data generated in Step~1,
      both using all $m$ variables in~(\ref{GLMaux2})
      and using only those $l$ in~(\ref{GLMaux}),
      obtaining new estimates $\doubletilde\beta$
      and $\doublehat\beta$, respectively, and
\item we calculate the test statistic
      (such as the discrete Kolmogorov-Smirnov distance $d$)
      using the new $y_1$,~$y_2$, \dots, $y_n$ generated in Step~1 and
      new estimates
      $\doublehat{\mu}_1$,~$\doublehat{\mu}_2$, \dots, $\doublehat{\mu}_n$,
      determining the ordering (that is, the permutation $\sigma$
      from Section~\ref{intro}) for the statistic by sorting
      $\doubletilde{\mu}_1$,~$\doubletilde{\mu}_2$, \dots,
      $\doubletilde{\mu}_n$, with
\begin{equation}
\label{GLMaux3}
\logit(\doublehat{\mu}_k) = \doublehat\beta^{(0)}
                          + \sum_{j=1}^l \doublehat\beta^{(j)} x_{j,k}
\end{equation}
and
\begin{equation}
\label{GLMaux4}
\logit(\doubletilde{\mu}_k) = \doubletilde\beta^{(0)}
                            + \sum_{j=1}^m \doubletilde\beta^{(j)} x_{j,k}
\end{equation}
      for $k = 1$,~$2$, \dots, $n$,
      where $\doubletilde\beta$ and $\doublehat\beta$ are the estimates
      calculated in Step~2 from the data generated in Step~1.
\end{enumerate}

After conducting many such simulations, we may estimate the P-value
as the fraction of the statistics calculated in Step~3
that are greater than or equal to the statistic calculated
from the given data.
The accuracy of the estimated P-value
is inversely proportional to the square root
of the number of simulations conducted, as detailed in the following remark.

\begin{remark}
\label{1error-bars}
The standard error of the Monte-Carlo estimate for an exact P-value $P$
is $\sqrt{P(1-P)/i}$, where $i$ is the number
of simulations conducted to produce the estimate.
Indeed, each simulation has probability $P$ of producing a statistic
that is greater than or equal to the statistic corresponding
to an exact P-value of $P$.
Since the simulations are all independent, the number of the $i$ simulations
that produce statistics greater than or equal to that corresponding
to P-value $P$ follows the binomial distribution
with $i$ trials and probability $P$ of success in each trial.
The standard deviation of the number of simulations whose statistics
are at least that corresponding to P-value $P$ is therefore
$\sqrt{i P (1-P)}$, and so the standard deviation
of the {\it fraction} producing such statistics
is $\sqrt{P (1-P)/i}$. Of course, the fraction itself
is the Monte-Carlo estimate of the exact P-value
(we can use this estimate in place of the unknown $P$
when calculating the standard error $\sqrt{P (1-P)/i}$).
\end{remark}

\section*{Acknowledgements}

We would like to thank Raymond Carroll and Peter McCullagh.
Mark Tygert was supported in part by an Alfred P.\ Sloan Research Fellowship
and a DARPA Young Faculty Award.
Rachel Ward was supported in part by an Alfred P.\ Sloan Research Fellowship,
a Donald D.\ Harrington Faculty Fellowship, ONR Grant N00014-12-1-0743,
an NSF CAREER Award, and an AFOSR Young Investigator Program Award.

\bibliographystyle{asamod.bst}
\bibliography{stat}

\end{document}